\documentclass{aa}  
\usepackage{graphicx}
\usepackage{txfonts}
\usepackage{float}
\usepackage{amsmath}
\usepackage{float}
\usepackage{natbib}
\begin{document}
   \title{Modelling total solar irradiance since 1878 from simulated magnetograms}

   \author{M. Dasi-Espuig\inst{1},
          J. Jiang\inst{2}
	  N. A. Krivova \inst{1},
          \and
          S. K. Solanki\inst{1,3}
          }

   \offprints{dasi@mps.mpg.de}

   \institute{Max-Planck-Institut f\"ur Sonnensystemforschung,
              Justus-von-Liebig-Weg 3, 37077 G\"ottingen, Germany\\
              \email{dasi@mps.mpg.de}
              \and
              {Key Laboratory of Solar Activity, National Astronomical Observatories, Chinese Academy of Sciences, Beijing 100012, China\\
             }
          \and
              School of Space Research, Kyung Hee University, Yongin, Gyeonggi, 446-701, Korea\\
             }
 
   \date{Received 27 May 2014 ; Accepted: 05 August 2014}

% \abstract{}{}{}{}{} 
% 5 {} token are mandatory
 
  \abstract
   {}
   {We present a new model of total solar irradiance (TSI) based on magnetograms simulated with a surface flux transport model (SFTM) and the SATIRE (Spectral And Total Irradiance REconstructions) model. Our model provides daily maps of the distribution of the photospheric field and the TSI starting from 1878. }
   {The modelling is done in two main steps. We first calculate the magnetic flux on the solar surface emerging in active and ephemeral regions. 
The evolution of the magnetic flux in active regions (sunspots and faculae) is computed using a surface flux transport model fed with the observed record of sunspot group areas and positions.
The magnetic flux in ephemeral regions is treated separately using the concept of overlapping cycles.
%Observational studies have shown that ephemeral regions follow a cyclic behaviour like that of sunspots, but their cycles are extended in time. To model the ephemeral region cycles, we assume that their length and amplitude are related to that of the sunspot cycles. Furthermore, since the ephemeral regions of a new cycle appear on the photosphere several years before the end of a cycle, this leads to an overlap that provides the secular variation of the total photospheric flux.
We then use a version of the SATIRE model to compute the TSI. The area coverage and the distribution of different magnetic features as a function of time, which are required by SATIRE, are extracted from the simulated magnetograms and the modelled ephemeral region magnetic flux. Previously computed intensity spectra of the various types of magnetic features are employed.
}
   {Our model reproduces the PMOD composite of TSI measurements starting from 1978 at daily and rotational timescales more accurately than the previous version of the SATIRE model computing TSI over this period of time. The simulated magnetograms provide a more realistic representation of the evolution of the magnetic field on the photosphere and also allow us to make use of information on the spatial distribution of the magnetic fields before the times when observed magnetograms were available.
We find that the secular increase in TSI since 1878 is fairly stable to modifications of the treatment of the ephemeral region magnetic flux.
%The relatively high TSI during the maxima of the early cycles (between 1878 and 1900) compared to the strength of these cycles in terms of sunspot numbers and areas may be a result of non-linear modulations in the evolution of the large-scale magnetic field introduced by the SFTM, since the SFTM includes the dependence of the cycle-averaged latitudes and tilts of active regions with the strength of a cycle.
 }
   {}


   \authorrunning{Dasi-Espuig et al.}
   \titlerunning{Modelling total solar irradiance since 1878}

   \maketitle
%
%________________________________________________________________

\section{Introduction}

Solar irradiance, which is the total energy flux from the Sun per unit area at 1 A.U., is the main external driver of the Earth's climate system, and therefore its variation is of interest when investigating the causes of global climate change.
Both correlation analyses \citep[e.g.][]{solanki03-b} and climate models \citep{haigh07, solomon07, jungclaus10-b, stocker13} have shown that solar irradiance variations alone cannot explain the observed sharp rise in the Earth's global temperatures after around 1970. Nonetheless, the exact role of the Sun in climate change remains uncertain, partly because the mechanisms through which the Sun affects climate are still under debate \citep{haigh01, haigh07, gray10, lockwood12, solanki13, stocker13}.
To properly asses such mechanisms, consistent long-term records of solar irradiance are needed. 
Multiple instruments on-board space satellites have been monitoring total solar irradiance (TSI) continuously, but only since 1978, i.e. over the last 3 and a half solar cycles \citep[e.g.][]{willson81, willson91, kopp05, frohlich03, frohlich09-b, kopp11}.
To study variations on timescales longer than the 11-year solar cycle, solar irradiance measurements need to be extended into the past with the use of models.

The first models of TSI variations based on proxies of solar magnetic activity (sunspots and faculae) appeared in the 1980s \citep{oster82, willson88, foukal88}.
Such models used linear regressions between the measured TSI and different proxies of solar activity.
The success of these models supported the assumption that irradiance changes on timescales longer than a day are due to the evolution of the magnetic fields on the solar surface and
motivated the development of more refined models \cite[e.g.][]{fligge00, krivova03, ermolli03, wenzler06, shapiro11, fontenla11}.
Amongst the most successful at present is the SATIRE-S (Spectral and Total Irradiance Reconstructions for the Satellite era) model, which captures more than 90\% of the observed variations in TSI \citep{krivova03, wenzler06, ball12} over the entire observational record \citep{yeo14}.
SATIRE-S uses spatially resolved, full-disc solar observations (magnetograms and continuum images) to track the evolving area coverage and distribution of faculae, network, and sunspots, together with semi-empirical model atmospheres from which the dependence of the brightness of each magnetic feature on the wavelength and its position on the disc is computed \citep{unruh99}.

Before 1974 high-resolution magnetograms that can be used in the SATIRE-S model are not available. Therefore, models based on proxies of the magnetic activity are needed. SATIRE-T  \citep{krivova07, krivova10} is a version of the SATIRE model that obtains the evolving area coverage of each magnetic feature from the sunspot number using a simple physical model of the evolution of the magnetic field on the solar surface \citep{solanki02-b, vieira10}. The sunspot number is a disc-integrated index of solar magnetic activity and thus the information on the spatial distribution of the magnetic features is lost. This means that the dependence of the brightness contrasts of individual surface features on the disc position cannot be taken into account and only disc-averaged contrast values can be employed. Thus, passages of individual active regions across the visible solar disc cannot be modelled as accurately as with SATIRE-S.

Over the last decade, surface flux transport models (SFTM) have proved to be a successful tool with which to describe the evolution of the large-scale magnetic field on the solar surface \citep{wang89, vanBallegooijen98, baumann04, schussler06}. Such models solve the radial component of the induction equation, where the active regions (sunspot and faculae) provide the source of the magnetic flux and the surface flows then determine the evolution of the large-scale field \citep[for more details about the models, see the reviews by][]{mackay12, jiang14-b}.
%Such models have successfully reproduced the major features of the sunspot cycle (such as the amplitude and duration), as well as the observationally inferred open flux since the beginning of the 20th century and the timing of the polar field reversals.
The SFTM can therefore provide the magnetic field on the solar surface at any instant in time, whenever we have measurements of sunspot group areas and positions.
A continuous record of sunspot group areas and positions with full coverage is available for the past 140 years, from the Greenwich photoheliographic maps between 1874 and 1976, and the USAF/NOAA SOON network after 1976.

\cite{wang05} reconstructed for the first time the TSI with the magnetic flux from an SFTM and the irradiance model of \cite{lean00}. They employed sunspot areas and numbers as well as the photospheric magnetic flux from an SFTM, to obtain the contribution to TSI from sunspots and faculae, respectively. However, their SFTM did not use the daily record of sunspot group positions and areas as input, but two different synthetic records based on the annual sunspot group number and thus only provided the TSI on a yearly basis.
In this paper we present a new model, SATIRE-T2 (where T2 refers to version 2 of the SATIRE-T) for reconstructing solar irradiance into the past on a daily basis using sunspot observations as input. For this, we adopt the SATIRE-S irradiance model to be fed by the simulated maps of the photospheric magnetic flux, which are produced with the version of the SFTM of \cite{cameron10}.
This SFTM takes into account the observed anti-correlation between the average sunspot group tilt angles and the strength of a cycle \citep{Dasi-Espuig10, dasi-espuig13, kitchatinov11, ivanov12, mcclintock13} as well as the correlation of the emergence latitudes of sunspot groups with cycle strength \citep{li03, solanki08, jiang11-a}. In particular, the dependency of the emergence latitudes on the cycle strength provides a saturation mechanism of the polar field and allows it to reverse even when a strong cycle is followed by a weak one.

Using the SFTM we simulated daily maps of the magnetic field on the solar surface, i.e. magnetograms, starting from 1874 (when the record of sunspot group areas and positions starts) to the present day.
The information on the spatial distribution of the magnetic features allowed us to take the centre to limb variation in the contrast of the surface magnetic features into account and to employ the SATIRE-S model for the reconstructions of TSI.
The SFTM, however, only includes the flux emerging in active regions since long-term and homogeneous records of ephemeral regions do not exist.
Therefore, we incorporated the flux in ephemeral regions in the same manner as was done in the earlier SATIRE-T model, i.e. based on the concept of overlapping cycles whose length and amplitude are related to those of the sunspot cycles.
Observational studies show that ephemeral regions follow the cyclic behaviour of sunspots, with cycles that start several years before the associated sunspot cycle \citep{harvey73, harvey93, harvey01, hagenaar03}.
Hence, the ephemeral regions are assumed to be part of the main dynamo process that produces active regions, and their cycle amplitude and length can be related to those of their corresponding sunspot cycles.
The extension in time of the ephemeral region cycles leads to an overlap of their cycles, and consequently, provides a secular change in the total magnetic flux \citep{solanki02-b, vieira10}.

Recent studies, both observational \citep[e.g.][]{lites08, ishikawa09, danilovic10, buehler13} and theoretical \citep{vogler07, schussler08}, hint at the existence of a local surface dynamo that could produce small bipolar regions \citep[for a review see][]{dewijn09}. However, the magnetic flux produced by such a dynamo is expected to be weaker than that traditionally considered for the ephemeral regions.

The paper is structured as follows. Section 2 describes the model used to obtain the photospheric magnetic flux in active regions and their decay products (with the SFTM), as well as in ephemeral regions.
In Sect. 3 we describe the fundamentals of the SATIRE-T2 model and how we adapted it to the simulated magnetograms. Section 4 presents the results of our model of TSI, and finally, in Sect. 5 we summarise our main conclusions.

%__________________________________________________________________

  \section{The photospheric magnetic flux}

   \subsection{Active region flux}

To describe the evolution of the magnetic flux on the solar surface that emerges in active regions we used an SFTM.
The model we used was originally developed by \cite{baumannthesis} and has been used and further developed by \cite{schussler06, cameron10, jiang10, jiang11}. 
It is a 2-D model that considers the transport of the large-scale magnetic field in the photosphere as a result of the effects of differential rotation, $\Omega$, meridional flow, $\nu$, and surface diffusivity, $\eta$, as a result of the random granular and supergranular motions. The model also assumes that the magnetic field in the photosphere is purely radial. This is reasonable since the magnetic field of faculae and network is only weakly inclined relative to the vertical \citep{solanki93, martinezpillet97}. Thus, the model solves the induction equation in spherical coordinates for the radial magnetic field component, $B$, as
\begin{eqnarray}
 \frac{\partial B}{\partial t} &=& -\Omega(\lambda)\frac{\partial B}{\partial \phi} - \frac{1}{R_{\odot}\cos(\lambda)}\frac{\partial}{\partial \lambda} \bigg(\nu(\lambda)B\cos(\lambda) \bigg) \nonumber \\
 &&+ \frac{\eta}{R^{2}_{\odot}} \bigg[\frac{1}{\cos\lambda}\frac{\partial}{\partial \lambda}\bigg(\cos\lambda\frac{\partial B}{\partial \lambda}\bigg)+\frac{1}{\cos^{2}\lambda}\frac{\partial^{2} B}{\partial \phi^{2}} \bigg]  \\
 &&+S(\lambda,\phi,t) \nonumber,
\end{eqnarray}
where $R_{\odot}$, $\lambda$, and $\phi$ are the solar radius, the latitude, and the longitude, respectively. 
The profiles of the differential rotation, $\Omega(\lambda)$, and the meridional flow, $\nu(\lambda)$, and the value of the surface diffusivity ($\eta=250$ km$^{2}$s$^{-1}$) used in this study are the same as those in \cite{jiang10} and \cite{cameron10}. The source term $S(\lambda,\phi,t)$ describes the emergence of new magnetic flux.
The radial magnetic field $B$ is expressed in terms of spherical harmonics. Following Baumann et al. (2004), we take the maximum order of the spherical harmonics to be $l$=63, which corresponds to a spatial resolution element roughly of the size of a supergranule (30 Mm).
 
The source function, $S(\lambda,\phi,t)$, dictates where, when, and with which size and tilt angle active regions appear on the solar surface.
The flux of a new active region is represented by two opposite polarity patches centred at $(\lambda_{+},\phi_{+})$ for the positive polarity, and $(\lambda_{-}, \phi_{-})$ for the negative polarity. The magnetic field distribution corresponding to each polarity is given by
\begin{eqnarray}
B^{\pm}(\lambda,\phi)=B_{max}\bigg({ \frac{0.4\Delta\beta}{\delta}}\bigg)^{2}\mathrm{exp}(-2[1-\cos(\beta_{\pm}(\lambda,\phi)]/\delta^2),
\end{eqnarray}
where $\beta_{\pm}(\lambda,\phi)$ are the angles between the centre of the positive/negative polarity patch and a point on the surface $(\lambda, \phi)$. 
The polarity separation, $\Delta\beta$, is the angular separation between the two polarities and is determined by the total area of the active region $A_{AR}$. We take $\Delta\beta=0.45A_{AR}^{0.5}$ following \cite{cameron10}, who determined this relationship using the Kodaikanal and Mount Wilson data sets.
The initial size (angular width of the Gaussian) of the individual polarity patches $\delta$ is $4^{\circ}$, which means that the smaller bipolar regions are considered only once they have already diffused to a width of $4^{\circ}$ \citep[see][for more details]{baumannthesis}, for numerical reasons.
$B_{max}$ is a scaling factor and was fixed to 374 G following \cite{cameron10} and \cite{jiang11}, who forced the total unsigned flux from the simulation results to match the measurements from the Mount Wilson and Wilcox Observatories.

In order to obtain the positions $(\lambda_{\pm},\phi_{\pm})$ of the two polarities of an active region, we need to know the group's central position, area, and the tilt angle $\gamma$ with respect to the azimuthal direction (see Eqs. (4) -- (7) in \cite{cameron10} for more details). The total area of an active region, $A_{AR}$, is expressed as the sum of the area of the sunspots, $A_{s}$, and the faculae, $A_{f}$.
The facular areas are estimated by using the observed relation from \cite{chapman97},
\begin{equation}
 A_{AR}=A_{s}+A_{f}=A_{s}+414+21A_{s}-0.0036A_{s}^{2},
\end{equation}
where areas are expressed in millionths of the solar hemisphere.
For sunspots, we used daily total sunspot group areas and positions from the data set compiled by the Royal Greenwich Observatory (RGO) from 1874 to 1976, and the USAF/NOAA Solar Optical Observing Network (SOON) from then on (hereafter referred to as the RGO-SOON data set)\footnote{http://solarscience.msfc.nasa.gov/greenwch.shtml}.
The SOON areas were multiplied by a correction factor of 1.4 to match the RGO ones, according to \cite{hathaway08} \citep[c.f.][]{balmaceda09}.
Since the SFTM is only responsible for the transport and decay of the magnetic flux, the growth phase of a group is not considered. Therefore, a group emerges in the SFTM on the day it reached its maximum area.

Neither the RGO nor SOON sunspot records contain information on the tilt angle of bipolar groups. The Mount Wilson and Kodaikanal Observatories provide the tilt angle data covering solar cycles 15 to 21.
In an analysis of the cycle-averaged tilt angles, \cite{Dasi-Espuig10, dasi-espuig13} found an anti-correlation between the mean normalised tilt angle of a given cycle and the strength of the same cycle. 
As in \cite{cameron10} and \cite{jiang11}, we considered the square root profile $\gamma=T_{n}\sqrt{|\lambda|}$ for each cycle $n$, where $\lambda$ and $\gamma$ are the latitude and the average tilt angle of sunspot groups, respectively, and $T_{n}$ is the constant of proportionality. Using this relationship \cite{jiang11} found a good anti-correlation between $T_{n}$ and the cycle strength. Here we extend the linear relationship between $T_{n}$ and the cycle strength throughout the whole simulation period to get the cycle-dependent tilt angles of sunspot groups, i.e. we assign all groups in one cycle the average tilt angle predicted by the linear relationship.
\begin{figure}
  \includegraphics[width=0.52\textwidth, trim={1.2cm 0 0 0.5cm}]{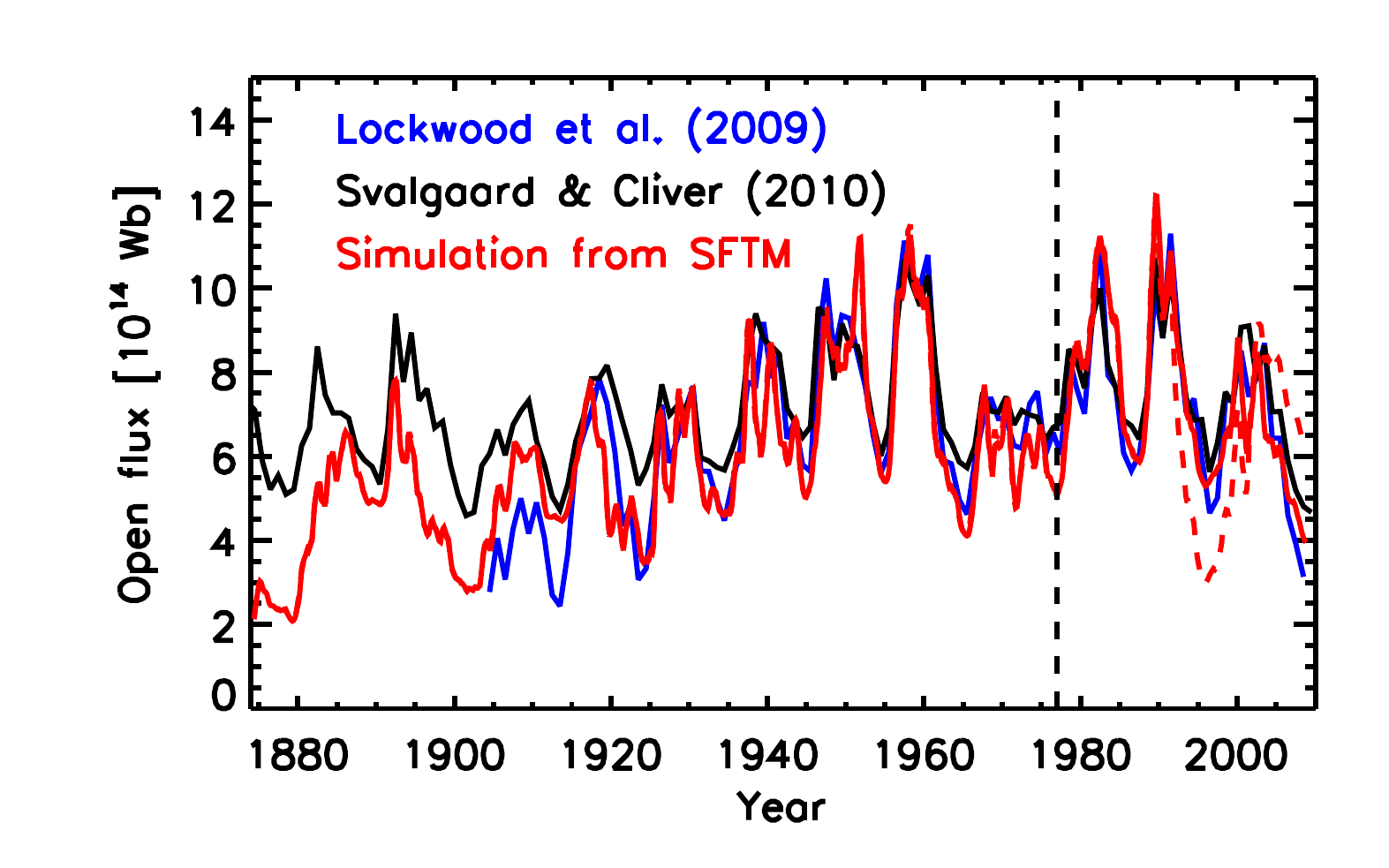}
  \caption{Open flux obtained with our SFTM before (red dashed curve) and after (red solid curve) an increase in the tilt angles in cycle 22 of 35\% (see text). The empirical reconstructions of the open flux by \cite{lockwood09} and \cite{svalgaard10} are shown by the blue and black curves, respectively.}
 \label{open_flux}
\end{figure}

The Sun's open magnetic flux is used to validate the SFTM. To obtain the open flux we took the simulated magnetograms and extrapolated the surface magnetic flux with the current sheet source surface model of \cite{zhao95, zhao95-b} following \cite{jiang10}.
In Fig.~\ref{open_flux} the simulated open flux (red dashed curve) is compared with two different empirical reconstructions based on geomagnetic indices:
the aa-index with kinematic correction \citep[][blue]{lockwood09}, and the InterDiurnal Variability index \citep[][black]{svalgaard10}. 
The vertical dashed line denotes the time when the sunspot group record switched from RGO to SOON.
Between 1920 and 1976, the simulated open flux matches both reconstructions fairly well. 
Before 1920 the two open flux reconstructions diverge, and our simulated results lie within the limits given by the two reconstructions.
We note that the open flux during the minima following the two weak cycles 12 and 14 (around 1890 and 1915, respectively) was comparatively high, whereas the stronger cycles that followed showed a low open flux at the end of the cycle (around 1900 and 1925).
This is due to the non-linear modulation of the dependence of the mean latitude and tilt angle with cycle strength, in the generation of the polar field \citep{cameron10, jiang13}.

After 1976, the two reconstructions of the open flux are in very good agreement, while the simulated open flux departs significantly from the two reconstructions, mainly during the minima after cycles 22 and 23 (around 1996 and 2009, respectively).
\cite{cameron12} argued that such disagreements can be due to the use of a constant factor to cross-calibrate the SOON and RGO areas \citep[see][]{balmaceda09, hathaway08}. The minimum measured area in RGO and SOON are different (1 and 10 millionths of a solar hemisphere, respectively) and therefore, a constant factor will overestimate the areas of the big groups \citep{foukal13}, which contribute by a disproportionate amount to the polar fields and the open flux.
\cite{yeates14} also showed that additional effects (e.g. changes in the meridional flow speed, in the sunspot group tilt angles, or in the value of the turbulent diffusivity) must be added to the standard flux transport model to simulate the evolution of cycle 23.
To circumvent this disagreement with the observations, here we artificially increased the tilt angles of cycle 22 by 35\% (red solid curve in Fig.~\ref{open_flux}). 
We note that the tilt angle records of Mount Wilson and Kodaikanal Observatories end around 1986, and it is unclear how this artificial increase compares to the real tilt angles in cycle 22.
The increase in the tilt angles caused a corresponding increase in the axial dipole field and the open flux by 35\% during the minimum of cycle 22, which in turn produced a decrease in the open flux during the minimum of cycle 23 of the same amount due an e-folding time of $\sim$4000 years of the large-scale magnetic field \citep{cameron10}.

  \begin{figure*}
   \centering
   \includegraphics[width=0.4\textwidth]{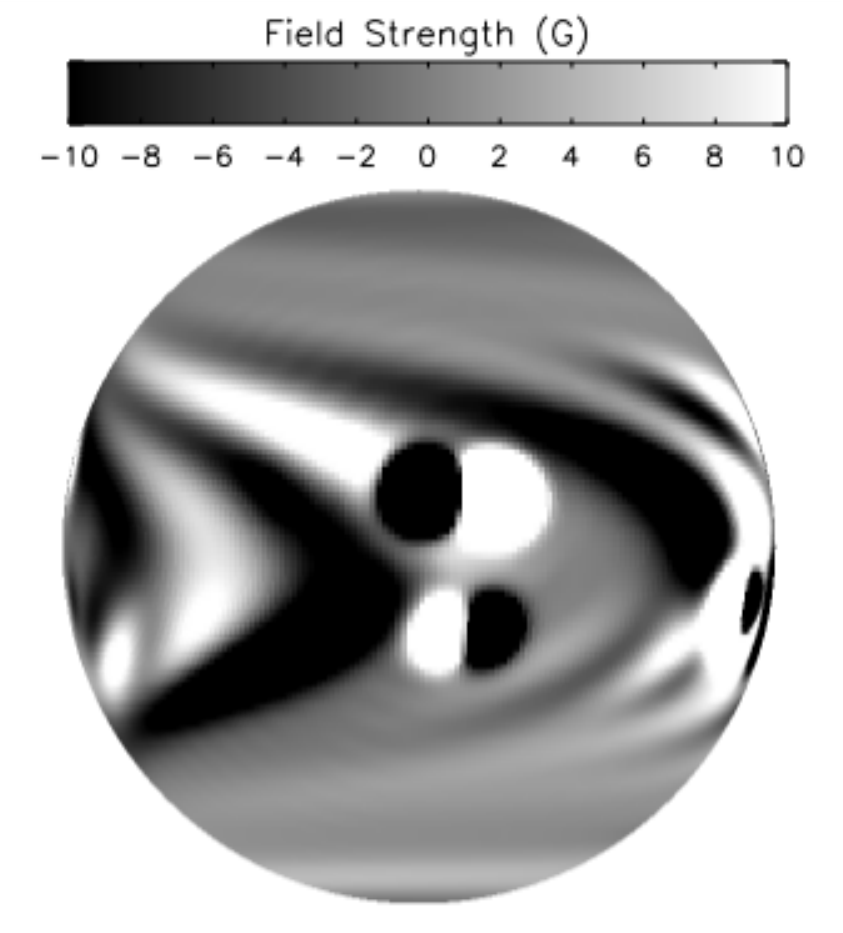}
   \hspace{7.00mm}
   \includegraphics[width=0.4\textwidth]{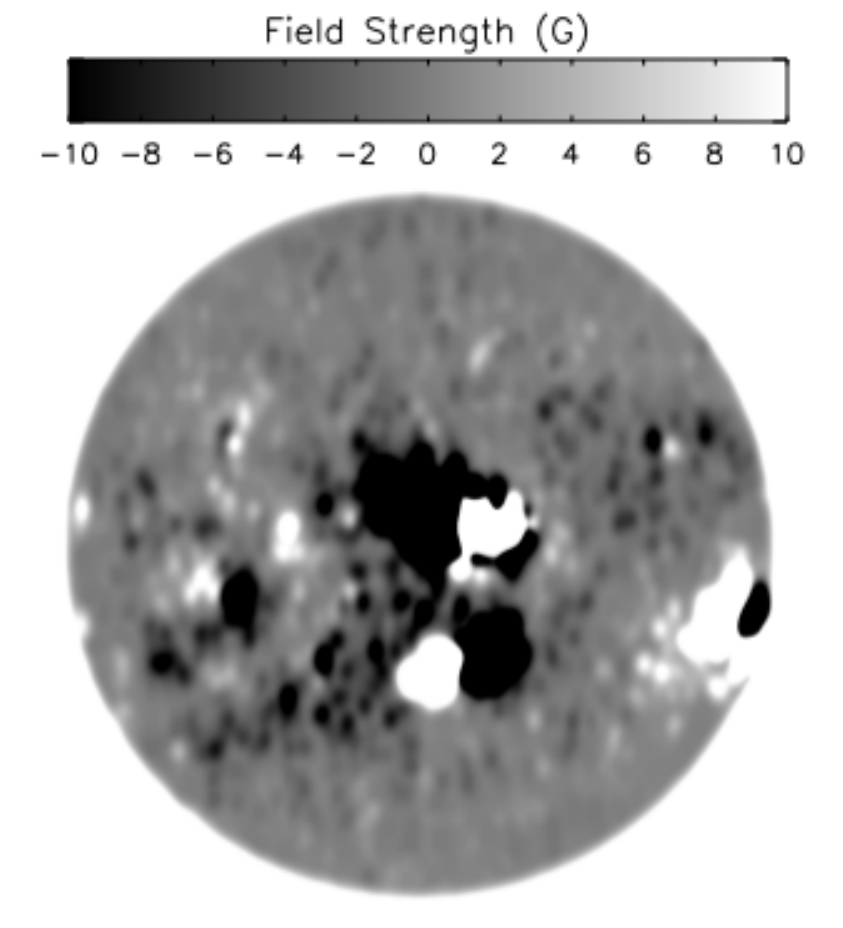}
      \caption{Simulated (left) and observed by MDI/SoHO (right) magnetogram on the 24th of July 2004. The observed magnetogram shows values above 3$\sigma$ of the noise level, corrected for the line-of-sight component to give the radial magnetic field strength. We then smoothed the observed magnetogram to resemble the simulated one more closely. The greyscale colour bars, which are the same for both images, indicate the field strength.}
      \label{magneto}
   \end{figure*}

 \subsection{Ephemeral region flux}

The SFTM does not account for the magnetic flux emerging in ephemeral regions (in the source term in Eq. (1)) mainly because there are no direct and continuous observations of ephemeral regions. \cite{harvey93, harvey01} found that the number of ephemeral regions present on the photosphere follows that of sunspots, but the duration of their cycles is extended in time. As a consequence, the magnetic flux from consecutive cycles overlaps, leading to a secular variation in the total photospheric flux \citep{solanki02-b}.
We therefore described the magnetic flux present in ephemeral regions as a function of time separately, as the sum of individual cycles whose length and amplitude are related to those of the corresponding sunspot cycles.

The total magnetic flux of ephemeral regions at a given time, $\Phi^{ER}(t)$, is
\begin{equation}
 \Phi^{ER}(t)=\sum_{n=1}^{N} \Phi_{n}^{ER}(t),
\end{equation}
where $\Phi_{n}^{ER}(t)$ is the magnetic flux of cycle $n$,
\begin{equation}
 \Phi_{n}^{ER}(t)= \Phi^{ER, max}_{n}g_{n}(t),
\end{equation}
and $N$ is the total number of cycles considered. The amplitude of cycle $n$, $\Phi^{ER, max}_{n}$, is chosen to vary linearly with the maximum sunspot group number of the corresponding cycle, $R_{n}^{max}$, relative to that of cycle 21, $R_{21}^{max}$. Following \cite{krivova07} and \cite{vieira10}, the amplitude of the magnetic flux in ephemeral regions in cycle 21, $\Phi^{ER, max}_{21}$, is then proportional to the amplitude of the magnetic flux in active regions in the same cycle, $\Phi^{AR, max}_{21}=60\cdot10^{14}$Wb \citep[estimated from Fig. 3 of][]{schrijver94}, through the scaling factor $X$,
\begin{equation}
 \Phi^{ER, max}_{n}=X\Phi^{AR, max}_{21}\frac{R_{n}^{max}}{R_{21}^{max}}.
\end{equation}
We also restrict the ratio between the minima and maxima of $\Phi^{ER}(t)$ in cycles 21 -- 23 to be equal or less than 2, consistent with observations \citep{harvey93, hagenaar03}.

The function $g_{n}(t)$ describes the dependence of $\Phi^{ER}(t)$ on time:
\begin{equation}
\begin{cases}
  g_{n}(t)=\cos^{2} \left( \frac{\pi(t-t_{n}^{m})}{L_{n}^{ER}} \right) & \rm{		} -L^{AR}_{n}/2-t_{x,n} \leq (t-t_{n}^{m}) \leq L^{AR}_{n}/2+t_{x,n} \\ 
 0 & \text{ otherwise.}
\end{cases}
\end{equation}
Here $t_{n}^{m}$ is the time at which the sunspot cycle $n$ reaches maximum activity and $L^{AR}_{n}$ is the length of the corresponding sunspot cycle, defined as the time between two consecutive minima. 
We use official dates for times of maxima and minima in the sunspot number\footnote{ftp://ftp.ngdc.noaa.gov/STP/space-weather/solar-data/solar-indices/sunspot-numbers/cycle-data}.
For cycle 24, which is still ongoing, we take the maximum of the smoothed monthly values until October 2013 and assigned it the average length of 11.1 years.
Although the smoothed sunspot number of cycle 24 peaked in October 2013\footnote{http://solarscience.msfc.nasa.gov/predict.shtml}, we do not know yet the exact length of cycle 24. This means that with the current data we are underestimating or overestimating the amount of ephemeral region flux during the declining phase of cycle 23 since the start of an ephemeral region cycle depends on the length of the associated sunspot cycle (see Eq. (7)).
The ephemeral region cycles are taken to peak at the time of sunspot cycle maxima and their length, $L_{n}^{ER}$, is extended with respect to the corresponding sunspot cycle by $2t_{x,n}$:
\begin{equation}
 L_{n}^{ER}=L_{n}^{AR}+2t_{x,n}.
\end{equation}
The time extension for cycle $n$, $t_{x,n}$, is related to the length of the $n$th sunspot cycle through a constant, $c_{x}$, so that longer sunspot cycles lead to larger extensions of ephemeral region cycles:
\begin{equation}
 t_{x,n}=L^{AR}_{n} - c_{x}.
\end{equation}

%The fact that we relate the length and amplitude of the ephemeral region cycle to the corresponding sunspot cycle implies that we are assuming that the ephemeral regions are produced by the global dynamo, like in the case of active regions.

 \subsection{Total flux}
 
The total magnetic flux is the sum of the magnetic flux in active regions and their decay products, and in ephemeral regions. As mentioned in Sect. 2.1, the active region flux calculated with the SFTM was scaled to match the measurements of the total magnetic flux from the Mount Wilson and Wilcox Solar Observatories with $B_{max}$=374 G \citep{cameron10, jiang11}. 
This did not take into account the contribution of the ephemeral region flux (not accounted by the SFTM) to the measured magnetic flux.
Therefore, we rescaled the magnetic flux from the SFTM, $\Phi_{AR}$, by a factor of $a\in[0,1]$, which is a free parameter in our model. Thus the final total flux is:
\begin{equation}
 \Phi^{tot}=a\Phi^{AR}+\Phi^{ER}.
\end{equation}

  \section{The solar irradiance model}
  
  \subsection{The SATIRE model}
  
The SATIRE model \citep[for a review see][]{krivova11-b} is based on the assumption that all irradiance variations are caused by the evolution of the magnetic field on the solar surface.
In particular, the SATIRE-S (S stands for the Satellite era) version of the model has proven to be very successful, being able to account for more than 90\% of the observed variations \citep{krivova03, wenzler05, ball12, yeo14}. SATIRE-S uses magnetograms to track the location and the extent of the magnetic features on the solar surface with time.

Here we extend the SATIRE-S model to times before the observed magnetograms were available, by simulating them using the SFTM (see Fig.~\ref{magneto}).
We divide the photosphere into 5 components: quiet Sun ($q$), sunspot umbra ($u$), sunspot penumbra ($p$), active region faculae ($f$), and network produced by ephemeral regions ($e$).
The irradiance contribution of each component is given by its intensity spectrum weighted by the fraction of the solar surface it covers.

The intensity spectra are independent of time, but depend on the wavelength, $\lambda$, and the heliocentric angle $\theta$ ($\mu=\cos\theta$).
They were calculated by \cite{unruh99} from the corresponding model atmospheres employing the ATLAS9 spectral synthesis code of \cite{kurucz92}. The sunspot umbrae and penumbrae are described by a radiative equilibrium model \citep{kurucz91} with effective temperatures of 4500 K and 5450 K, respectively; the quiet Sun by the standard model atmosphere FAL-C \citep{fontenla93} with an effective temperature of 5777 K; and both the faculae and the network by a version of FAL-P \citep{fontenla93} modified by \cite{unruh99}.

We introduce filling factors, $\alpha^{c}(\mu, t)$, that describe the fraction of the solar surface lying in the range $[\mu - \Delta \mu, \mu + \Delta \mu)$, where $\Delta \mu << 1$, that is filled by component $c$ at a given instant in time $t$.
They are derived as in the SATIRE-S model, but employing the simulated magnetograms instead of observed ones. 
The SFTM provides a daily simulated magnetogram, i.e. a map of the radial magnetic field of the whole solar surface (360$^{\circ}$ in longitude and 180$^{\circ}$ in latitude).
With a pixel size of 1$^{\circ}$x1$^{\circ}$, a single pixel is big enough to cover multiple types of the magnetic features distinguished by the SATIRE-S model, namely faculae, sunspot umbra, and sunspot penumbra.
In order to calculate the fraction of a pixel covered by each component we first resampled the magnetic field maps covering the solar surface by dividing each pixel $(i,j)$ into 100 sub-pixels.
The resampled maps of the radial field were then projected to the Sun's visible hemisphere to obtain the projected area contribution of each pixel facing the Earth.
Such a magnetogram is displayed in Fig.~\ref{magneto} together with an MDI magnetogram on the same day. For a direct comparison between the two magnetograms, we removed the pixels in the MDI magnetogram below 3$\sigma$ of the noise, and divided the magnetogram signal in each pixel by its $\mu$ value to obtain the radial component of the field.
We note that our simulated magnetograms are free of noise, and therefore every pixel is used to compute filling factors.
Finally, we computed the filling factors as a function of $\mu(k,l)$, where $(k,l)$ represent the pixels of the resampled and projected maps, as described below.

The sunspot group areas and positions for the TSI reconstruction were taken from the RGO-SOON daily data set, consistent with the data set used in the SFTM. Furthermore, since the SFTM only considers the decaying phase of the active regions, we used the areas and positions starting from the day the groups reached maximum area. 
For each simulated magnetogram we identified the pixels $(k,l)$ that correspond to the central position of the observed sunspot groups on that day, as well as their area coverage. Since the RGO-SOON data set does not list the areas and positions of the individual spots within a group, we assumed that a group is composed of two circular portions with half the size of the total group area, and masked the corresponding pixels.
The sunspot filling factor for each pixel $(k,l)$ is thus equal to 1.
We then computed the number of sunspot pixels as a function of $\mu$ for each day, $\alpha^{s}(\mu, t)$, in bins of $\Delta\mu=0.01^{\circ}$.
To separate the umbral and penumbral contributions we used the fixed ratio of umbral to total sunspot area, $A_{u}/A_{s}=0.2$ \citep{brandt90, wenzler05}, with $A_{s}=A_{u}+A_{p}$. Therefore each $\mu$ bin is assigned a value of $\alpha^{u}(\mu,t)=0.2 \alpha^{s}(\mu,t)$ and $\alpha^{p}(\mu,t)=0.8 \alpha^{s}(\mu,t)$ for sunspot umbra and penumbra, respectively.
%The sunspot filling factors, $\alpha^{s}_{\mu(k,l)}$, are then given by the sum of all pixels identified as part of a sunspot group. 
%To separate the umbral and penumbral contributions we used the fixed ratio of umbral to total sunspot area, $A_{u}/A_{s}=0.2$ \citep{brandt90, wenzler05}, with $A_{s}=A_{u}+A_{p}$. 
%This way $\alpha^{u}_{\mu(k,l)}=0.2\alpha^{s}_{\mu(k,l)}$, and $\alpha^{p}_{\mu(k,l)}=0.8\alpha^{s}_{\mu(k,l)}$.

All other pixels that are not part of a sunspot group are considered as mixed regions of faculae and quiet Sun.
For such pixels we computed their number as a function of $\mu$ (in bins of $\Delta\mu=0.01^{\circ}$) and the magnetic field strength (in bins of $\Delta B=2$ G) for every day: $H(\mu, B, t)$.
The facular filling factors for every $B$ bin are then calculated as in the SATIRE-S model \citep[e.g.][]{krivova03, wenzler04, ball12}, i.e. they increase linearly with the magnetic field strength, reaching unity at a saturation value $B^{f}_{sat}$. This saturation takes into account that magnetic features in weaker faculae are brighter per unit magnetic flux \citep{solanki84,foukal84,ortiz02}:
\begin{equation}
\alpha^{f}(\mu, B, t) = \left\{ 
\begin{array}{l l}
  H(\mu, B, t) \cdot B/B^{f}_{sat} & \quad \mbox{if $B$ < $B^{f}_{sat}$}\\
  H(\mu, B, t) \cdot 1 & \quad \mbox{if $B$ $\geq$ $B^{f}_{sat}$}.\\ \end{array} \right.
\end{equation}
Here $B^{f}_{sat}$ is a free parameter of the model. The sum over the $B$ bins in $\alpha^{f}(\mu, B, t)$ provides the total fraction of pixels filled by faculae for each narrow band of $\mu$: $\alpha^{f}(\mu, t)=\sum_{B} \alpha^{f}(\mu, B, t)$.
The part of each pixel which is not covered by spots or faculae is considered to be the quiet Sun, and the remaining fraction in each $\mu$ bin is
\begin{equation}
\alpha^{q}(\mu, t) = N(\mu) -\alpha^{u}(\mu, t)-\alpha^{p}(\mu, t)-\alpha^{f}(\mu, t),
\end{equation}
where $N(\mu)$ is the total number of pixels.
The filling factors of the ephemeral regions, $f^{e}$, are disc integrated since their magnetic flux was calculated for the full disc from Eqs. (4) -- (9).
We proceeded in the same way as for the facular filling factors, i.e. $f^{e}$ increases linearly until a saturation value, $B^{e}_{sat}$ is reached.

Finally, the irradiance was calculated in SATIRE-T2 as
\begin{eqnarray}
  S(t,\lambda)&=&\sum_{\mu} \bigg[ 
    \alpha^{u}(\mu,t) I^{u}(\mu,\lambda) 
 + \alpha^{p}(\mu,t) I^{p}(\mu,\lambda) \nonumber \\
&& + \alpha^{f}(\mu,t) I^{f}(\mu,\lambda)
 + \alpha^{q}(\mu,t) I^{q}(\mu,\lambda)
    \bigg] \\
 & & \nonumber \\
 &&+ f^{e}(t) F^{f}(\lambda) - f^{e}(t)F^{q}(\lambda) \nonumber,
\end{eqnarray}
where the quantities $F^{f}(\lambda)$ and $F^{q}(\lambda)$ are the disc integrated intensities of faculae and the quiet Sun, respectively. The last term subtracts the fraction of the solar surface covered by the quiet Sun that has been replaced by ephemeral regions. Integrating $S(t, \lambda)$ over all wavelengths provides the TSI.

 \subsection{Parameters and optimisation}
\begin{table}[t]
\begin{minipage}{\columnwidth}
 \caption{Free parameters of the model that provide the best fit to the total photospheric flux and the total solar irradiance, and their allowed ranges.}
 \centering
 \small
 \renewcommand{\footnoterule}{}  % to avoid a line before footnotes
 \begin{tabular}{l c c c }
 \hline\
   Parameter  &  Notation  &  Value  & Range\\
 \hline
 AR\footnote{Active Region} scaling factor			& $a$ & 0.68 & 0 -- 1 \\ 
 ER\footnote{Ephemeral Region} amplitude factor 		&  $X$ & 0.62 & 0.5 -- 1.2 \\
 ER cycle extension [years] 	&  $c_{x}$          & 7.45 & 5 -- 9 \\
 Saturation flux of faculae [G] 	& $B_{sat}^{f}$  & 513 & 100 -- 600 \\
 Saturation flux of ER [G] 		& $B_{sat}^{e}$ & 813 & 500 -- 1200 \\
 \hline
 \end{tabular}
 \end{minipage}
\end{table}

Our model has 5 free parameters in total: the scaling factor of the total simulated flux, $a$ (Sect. 2.3); the amplitude, $X$, and the time extension, $c_{x}$, of the ephemeral region cycle with respect to the sunspot cycle (Sect. 2.2); as well as the saturation values for the faculae, $B^{f}_{sat}$, and the ephemeral regions, $B^{e}_{sat}$ (Sect. 3.1).
Other parameters entering the SFTM (profiles of the differential rotation and meridional flow, surface diffusivity, polarity separation of sunspot groups, sunspot to facula area ratio, and tilt angles) were taken from observations and fixed for the full period of reconstructions (Sect. 2.1).

We employed the optimisation routine PIKAIA\footnote{http://www.hao.ucar.edu/modeling/pikaia/pikaia.php} \citep{charbonneau95} to find the values of the free parameters that minimise the combined reduced chi square, $\chi^{2}_{r}$, of the modelled and observed total magnetic flux (TF) and total solar irradiance (TSI): $\chi^{2}_{r,TF} + \chi^{2}_{r,TSI}$.
We defined the reduced chi square as
\begin{equation}
 \chi^{2}_{r} = \frac{1}{\nu}\sum_{i=1}^{P} \left( \frac{x^{obs}_{i}-x^{model}_{i}}{\sigma^{obs}} \right)^{2},
\end{equation}
where $\nu$ is the degrees of freedom, and P is the total number of observed data points. 
$x^{obs}_{i}$ and $x^{model}_{i}$ correspond to the observed and modelled data points, respectively. Finally, we weight each data point by the standard deviation of the measured time series, $\sigma^{obs}$, since individual errors are unknown.
 \begin{figure*}
  \includegraphics[width=0.51\textwidth]{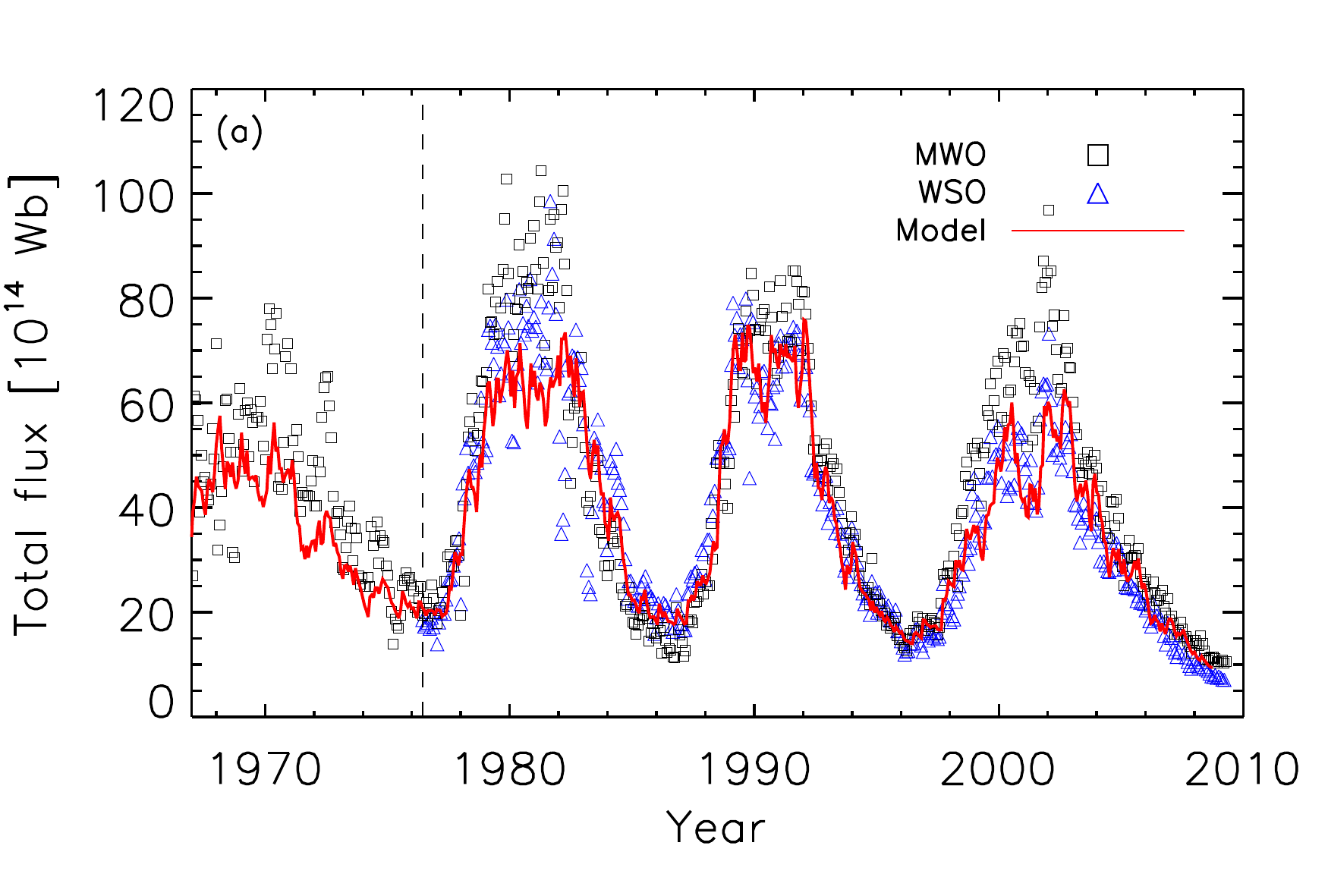}
  \includegraphics[width=0.51\textwidth]{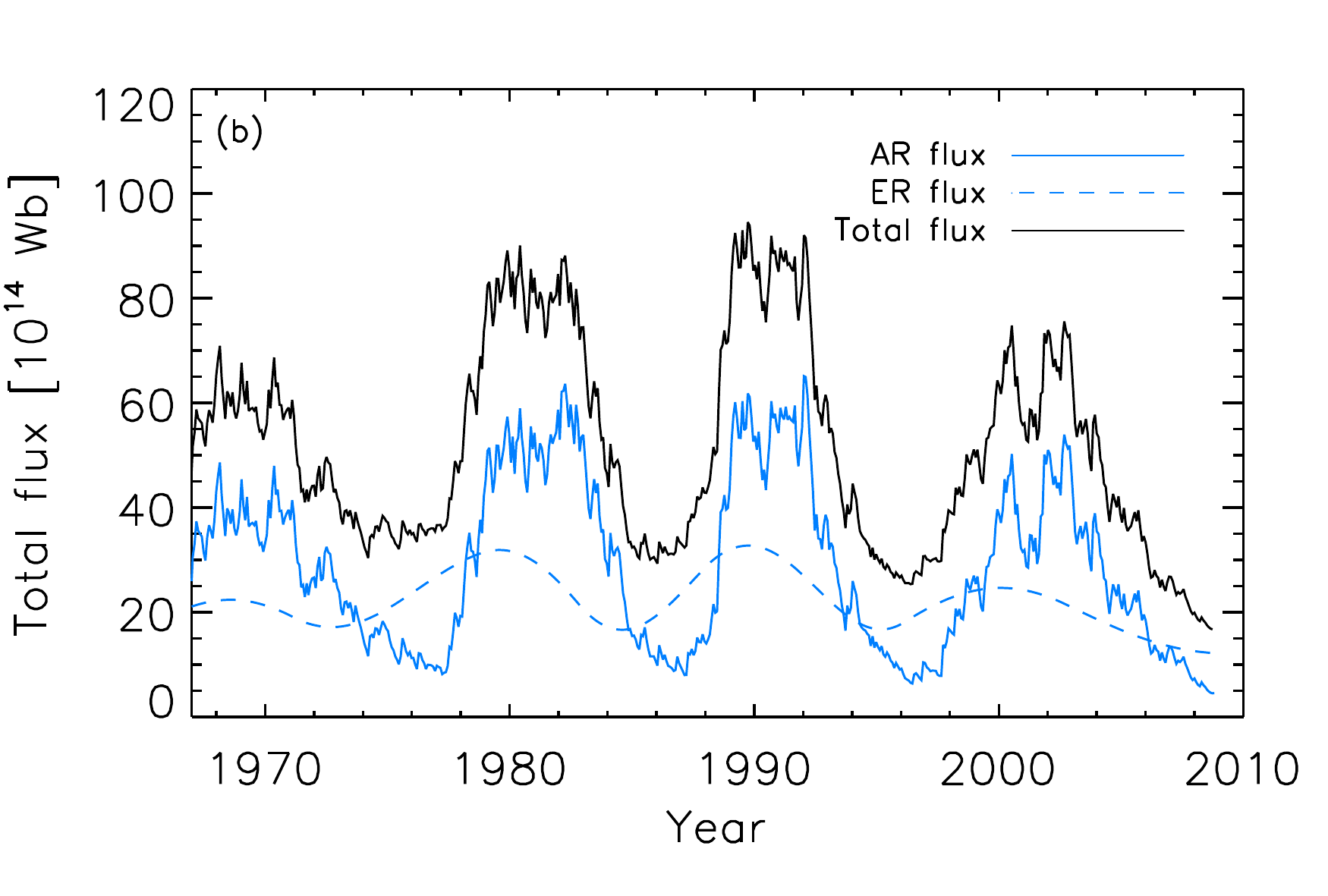}
  \caption{a) Modelled total magnetic flux (solid red), $0.68\Phi_{AR}+0.4\Phi_{ER}$ (see Sect. 3.3), compared with the observations from the Wilcox Solar Observatory (blue triangles) and the Mount Wilson Observatory (black squares). The vertical dashed line indicates the time when the data from the Mount Wilson observatory become available.
  b) Modelled magnetic flux from active regions ($0.68\Phi_{AR}$; solid blue), ephemeral regions ($\Phi_{ER}$; dashed blue), and the total magnetic flux ($0.68\Phi_{AR}+\Phi_{ER}$; solid black).}
 \label{totf}
\end{figure*}

The total magnetic flux measurements from the Mount Wilson and Wilcox Solar Observatories are available since 1967 and 1976, respectively \citep{arge02,wang06}. For the comparison with the modelled total magnetic flux we used the data during the period of overlap between the two observatories, i.e. between 1976 and 2009.
To take into account that at least half of the magnetic flux emerging in ephemeral regions goes undetected because of the insufficient spatial resolution of these relatively low-resolution magnetograms \citep{krivova04} we multiplied the magnetic flux in ephemeral regions by $0.4$ when comparing to the observations, following \cite{krivova07}.
The total magnetic flux was finally calculated as  $a\Phi^{AR}+0.4\Phi^{ER}$.
We note that the open flux here is included in $\Phi_{AR}$, in contrast to the model by \cite{vieira10}.
The modelled TSI was compared to the composite of TSI measurements provided by the Physikalisch-Meterologisches Observatorium Davos (PMOD) since 1978 \citep[][version d41\_62\_1302]{frohlich05}.
This is the TSI composite that agrees by far the best with the TSI models \citep{wenzler09, yeo14}.

Table 1 lists the values of the free parameters that minimise Eq. (14) and the limits within which we allowed them to vary. 
The scaling factor $a$ was allowed to vary between 0 and 1, because the SFTM alone is set to reproduce the observations of total magnetic flux with $B_{max}=374$ G and $a=1$ without the addition of ephemeral region flux. Their introduction implies a lower value of $a$. We found that a value of $a=0.68$ provides the best fit to the observed total magnetic flux (Fig.~\ref{totf}).
We stress that the parameter $B_{max}$ used in the SFTM \citep[e.g.][]{cameron10} is not related to the saturation values $B^{f}_{sat}$ and $B^{er}_{sat}$. While $B_{max}$ is used solely by the SFTM to scale the magnetic flux in active regions and their decay products to the observations (see Eq. (2)), the saturation values are employed to compute the filling factors, and thus the irradiance (see Eq. (13)).

The time extension, $c_{x}$, and the amplitude factor, $X$, of the ephemeral region cycles were constrained following \cite{krivova07, krivova10} and \cite{vieira10}.
For the amplitude factor $X$ we chose the range between 0.5 and 1.2.
This range is effectively identical to the one employed by \cite{krivova07, krivova10} and \cite{vieira10}, taking into account the different units employed in each study.
Our best-fit values are $X=0.62$ and $c_{x}=7.45$. With $c_{x}=7.45$ the ephemeral region cycles start around 3 to 4 years before active region minima, and have an average length of 18 years. These values are different from those obtained by \cite{krivova10} owing to the different models used to describe the evolution of the photospheric flux emerging in active regions.

The parameter $B^{f}_{sat}$ represents the value of the magnetic flux per magnetogram pixel above which the pixel is considered to be completely filled by the facular component. The saturation value depends on the spatial resolution, noise level and sensitivity of the magnetograms employed.
Since the simulated magnetograms are noise-free, the value of $B_{sat}^{f}$ is expected to be higher here as compared to the observed magnetograms \citep[see][]{krivova07}.
%The SATIRE-S does not distinguish between faculae and network as done here and in \cite{krivova07, krivova10}, and uses only one saturation value to fit the TSI. 
The range of values previously found with the SATIRE-S is 200 G -- 443 G, when employing magnetograms from the Kitt Peak Vacuum Telescope, MDI/SoHO, or HMI/SDO \citep{krivova03, wenzler05, wenzler06, ball12, yeo14}.
Thus, we allowed $B_{sat}^{f}$ to vary between 100 G and 600 G.
For the saturation value of ephemeral regions, $B^{er}_{sat}$, we chose a range between 500 G and 1200 G, centred around the value estimated previously by \cite{krivova10}.

%We limited the saturation value of faculae, $B^{f}_{sat}$, to be within 100 G and 600 G, based on the range of values  found previously with the SATIRE-S \citep[200 G -- 443 G;][]{krivova03, wenzler05, wenzler06, ball12, yeo14}.
%\textbf{The value of $B^{f}_{sat}$ represents the value of the magnetic field strength of a pixel in a magnetogram above which the region inside that pixel is considered to be completely filled by our facular atmosphere. Therefore, the saturation values will be different depending on the spatial resolution, noise level, and sensitivity of the magnetograms employed.
%As done in the  here we use separate saturation values for faculae and ephemeral regions}
%\textbf{For the saturation value of ephemeral regions, $B^{e}_{sat}$, we chose a range between 500 G and 1200 G, centred around the value estimated previously by \cite{krivova10}.}
%As discussed by \cite{krivova10}, we expected the saturation value of ephemeral regions, $B^{e}_{sat}$, to be around 800 G. We therefore chose a range centred around this estimate, between 500 G and 1200 G.

%______________________________________________________________ 
 \section{Results}

 \subsection{Total solar irradiance since 1978}

%TOTAL FLUX --  Fig. 3
In this section we compare the output of our model over the period of satellite observations to the measured total magnetic flux and TSI.
Figure~\ref{totf}a compares the modelled (red) photospheric magnetic flux to the measurements by the Wilcox Solar Observatory (blue triangles) and the Mount Wilson Observatory (black squares) over the past 3.5 solar cycles. We note that for the comparison with observations we multiplied the ephemeral region flux by 0.4 to account for the fraction of the flux that goes undetected in the magnetograms because of the limited spatial resolution \citep{krivova04}. The correlation coefficient between the model and the average of the two data sets is $r_{c}=0.95$, while the slope of the linear regression is $s=1.11$.
These values were determined for the period after 1976, when both data sets were available (i.e. to the right of the vertical dashed line in Fig.~\ref{totf}a).
Figure~\ref{totf}b depicts the total magnetic flux (solid black) and the contribution from active (solid blue) and ephemeral (dashed blue) regions over the same period of time (without multiplying by the factor of 0.4).

\begin{table}[b]
\begin{minipage}{\columnwidth}
 \caption{Slope, correlation coefficient ($r_{c}$), and reduced chi squared ($\chi_{r}^{2}$) between the modelled and observed total magnetic flux and TSI for different averaging periods.}
 \centering
 \small
 \renewcommand{\footnoterule}{}  % to avoid a line before footnotes
 \begin{tabular}{l l c c c }
 \hline
   Quantity  &  Scale  &  Slope  & $r_{c}$  &  $\chi_{r}^{2}$ \\
 \hline
 Total magnetic flux	& 1 CR\footnote{Carrington Rotation}
           						& 1.11 & 0.95 & 0.124  \\ 
 TSI				& 1 day 		& 0.83 & 0.78 & 0.413  \\
 TSI				& 3 months	& 0.96 & 0.90 & 0.189  \\
 \hline
 \end{tabular}
 \end{minipage}
\end{table}

\begin{figure*}
   \centering
    \includegraphics[width=\textwidth]{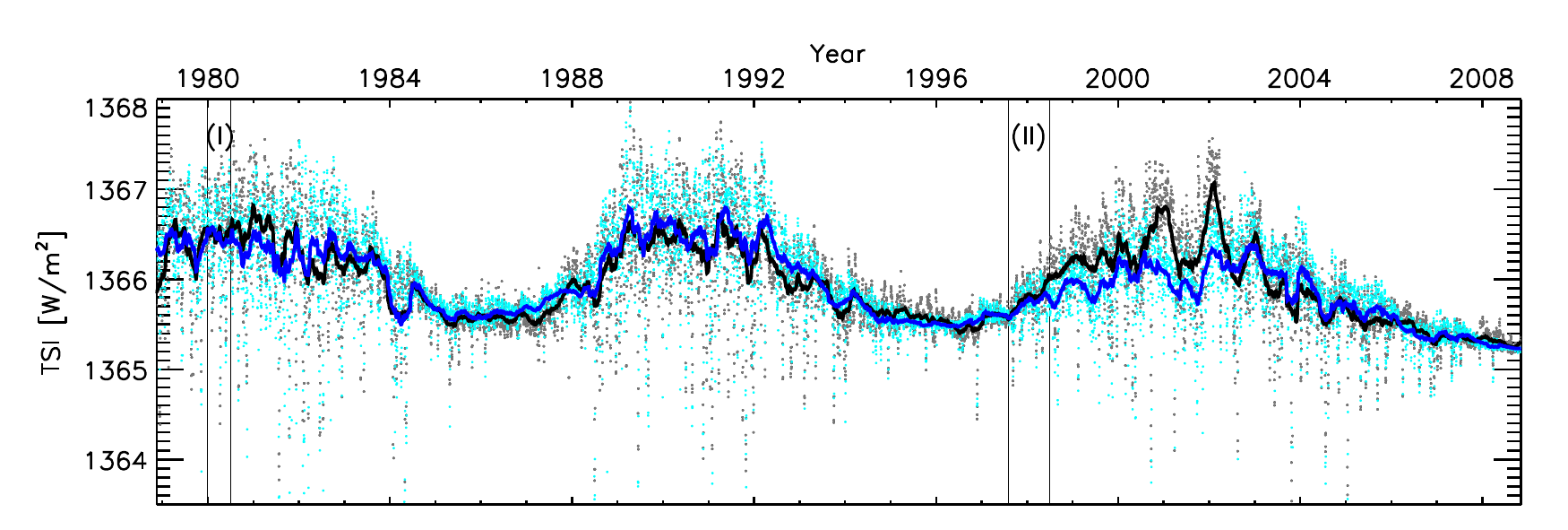}
    
    \includegraphics[width=\textwidth]{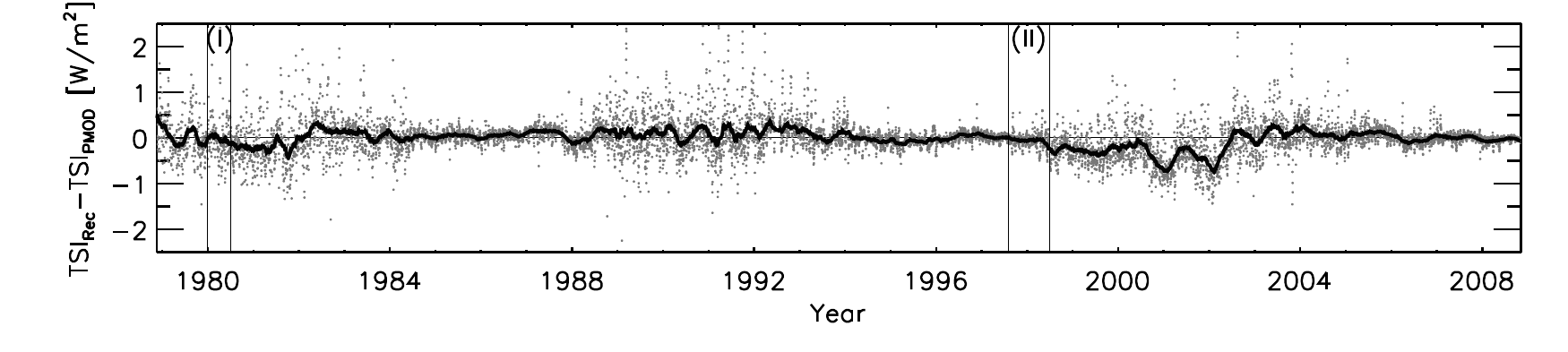}
      \caption{Upper panel: TSI reconstruction (blue) and the PMOD composite of observations (black). The dots represent daily values and the thick solid lines 3-month running means. The vertical lines mark two shorter periods shown in Fig.~\ref{TSI_month}. Lower panel: the difference between the reconstruction and the PMOD composite. }
      \label{diff_er}
\end{figure*}

%TSI -- Fig. 4 -- general
The modelled TSI since 1978 (blue) is compared with the PMOD composite of observations (black) in the upper panel of Fig.~\ref{diff_er}. The lower panel displays the difference between the two. In both panels each point corresponds to a daily value, while the thick solid lines are 90-day running means. 
The overall shape and the amplitude of the TSI cycles are reproduced adequately by SATIRE-T2. In particular, the rising and declining phases are reproduced better than with the SATIRE-T \citep{krivova07, krivova10}.
The correlation coefficient, the slope, and the reduced $\chi^{2}$ as defined in Eq. (14) are given in Table 2, both for the daily values and the 90-day running means of the TSI. 
%In this case we did not weight the data points by the standard deviation of the measured time series as done in Eq. (14), to enable a direct comparison with the $\chi^{2}$ values given in \cite{krivova10}.

\begin{figure*}
   \centering
     \includegraphics[width=0.49\textwidth]{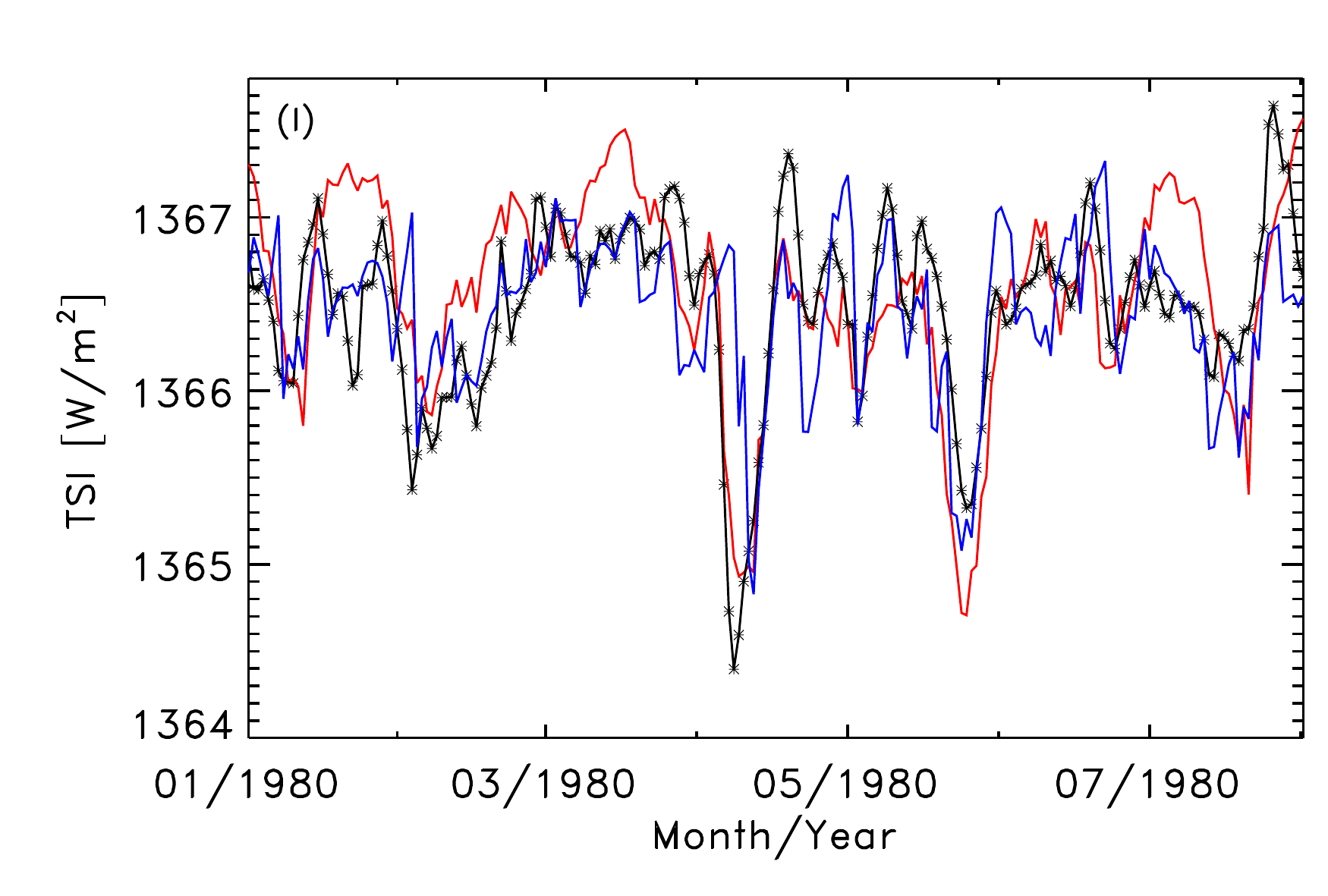}
     \includegraphics[width=0.49\textwidth]{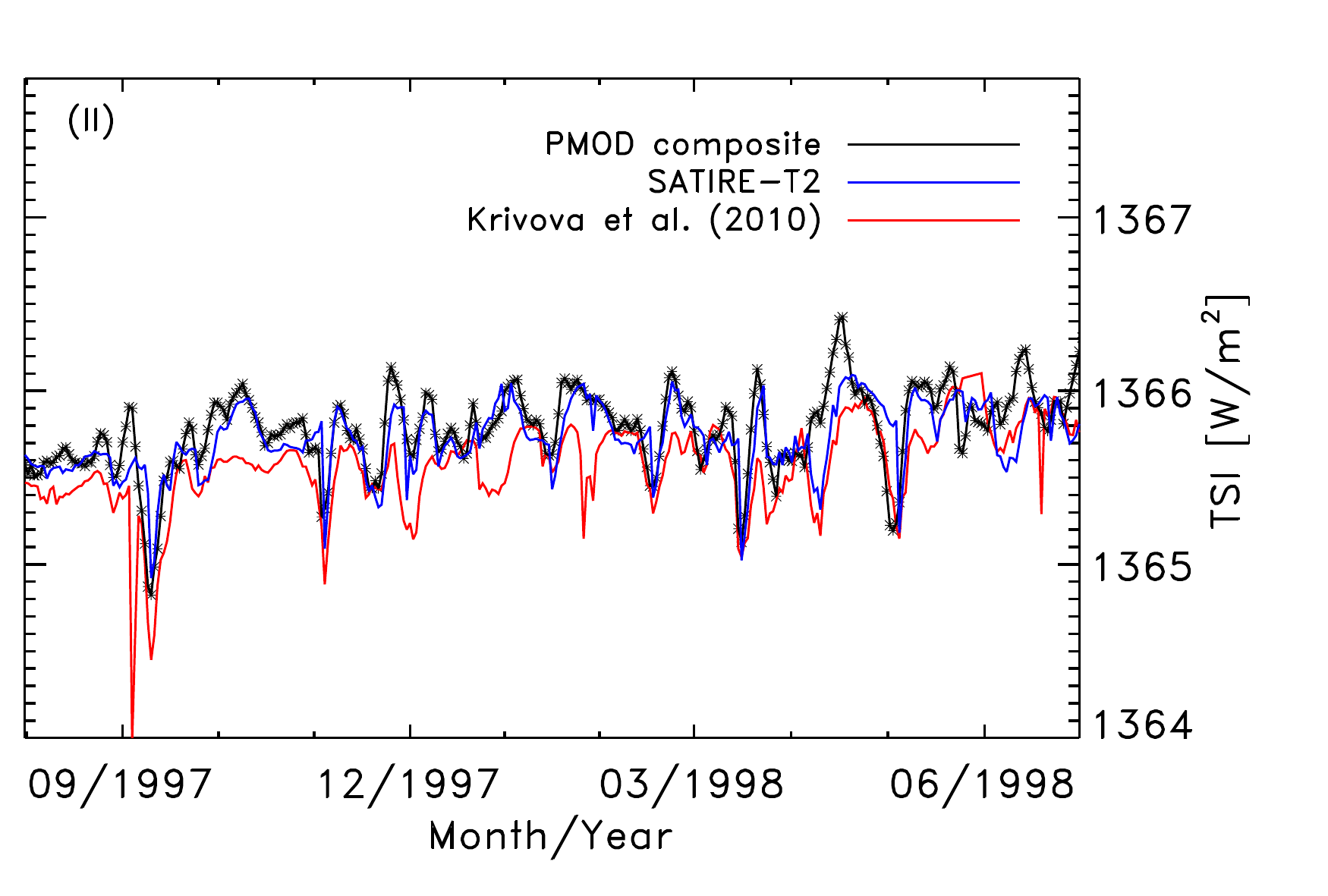}

      \caption{Enlargements of two shorter periods corresponding to the sections labelled (I) and (II) between the vertical lines in Fig.~\ref{diff_er}. Black, blue, and red represent the PMOD composite of TSI observations, our reconstruction, and the reconstruction of \cite{krivova10}, respectively.}
      \label{TSI_month}
\end{figure*}

%TSI -- Fig. 5 -- daily/rotational time scales
Figure~\ref{TSI_month} is an enlargement of two shorter periods of 8 and 10 months, each marked in Fig.~\ref{diff_er} by the vertical lines and the roman numerals (I) and (II). 
For comparison, we included the reconstruction from the SATIRE-T
\footnote{http://www2.mps.mpg.de/projects/sun-climate/data.html} model \citep{krivova10}, indicated by the red curve in Fig.~\ref{TSI_month}.
The new model follows changes on daily and rotational timescales better than the old version of SATIRE-T because of a more realistic description of the evolution of the facular fields by the SFTM and the employment of spatially resolved quantities instead of disc-integrated ones for computing the TSI.
Since intensity contrasts of the various magnetic structures depend strongly on the viewing angle \citep[e.g.][]{topka97, yeo13}, information on the spatial distribution of the magnetic features provided by the simulated magnetograms helps to improve the quality of the irradiance reconstruction, in particular on short timescales.
The model of the evolution of the photospheric magnetic flux in SATIRE-T only provided disc-integrated quantities.

%TSI -- details on daily variations
When looking in detail one can see that individual dips in irradiance caused by passages of spots across the disc are often narrower and start slightly later in the reconstruction than in the observations. An example is the big dip in panel (I) of Fig.~\ref{TSI_month} in April 1980 because the SFTM and thus SATIRE-T2 only consider sunspot groups from the time they reach their maximum area onwards (see Sect. 2.1), and therefore miss the contribution of sunspot groups during their growth phase.
%This may also affect the faculae. 
%Assuming that no cancellation of magnetic flux occurs during the growth phase of an active region, the total magnetic flux emerged during the active region's life-time can be approximated to the magnetic flux at the time it reaches maximum area. 
%However, during times of cycle maxima this approximation is not completely valid since many active regions populate the solar surface close to each other and cancellation of magnetic flux can occur between neighbouring regions before they have reached maximum area. 

\begin{figure*}
  \includegraphics[width=\textwidth]{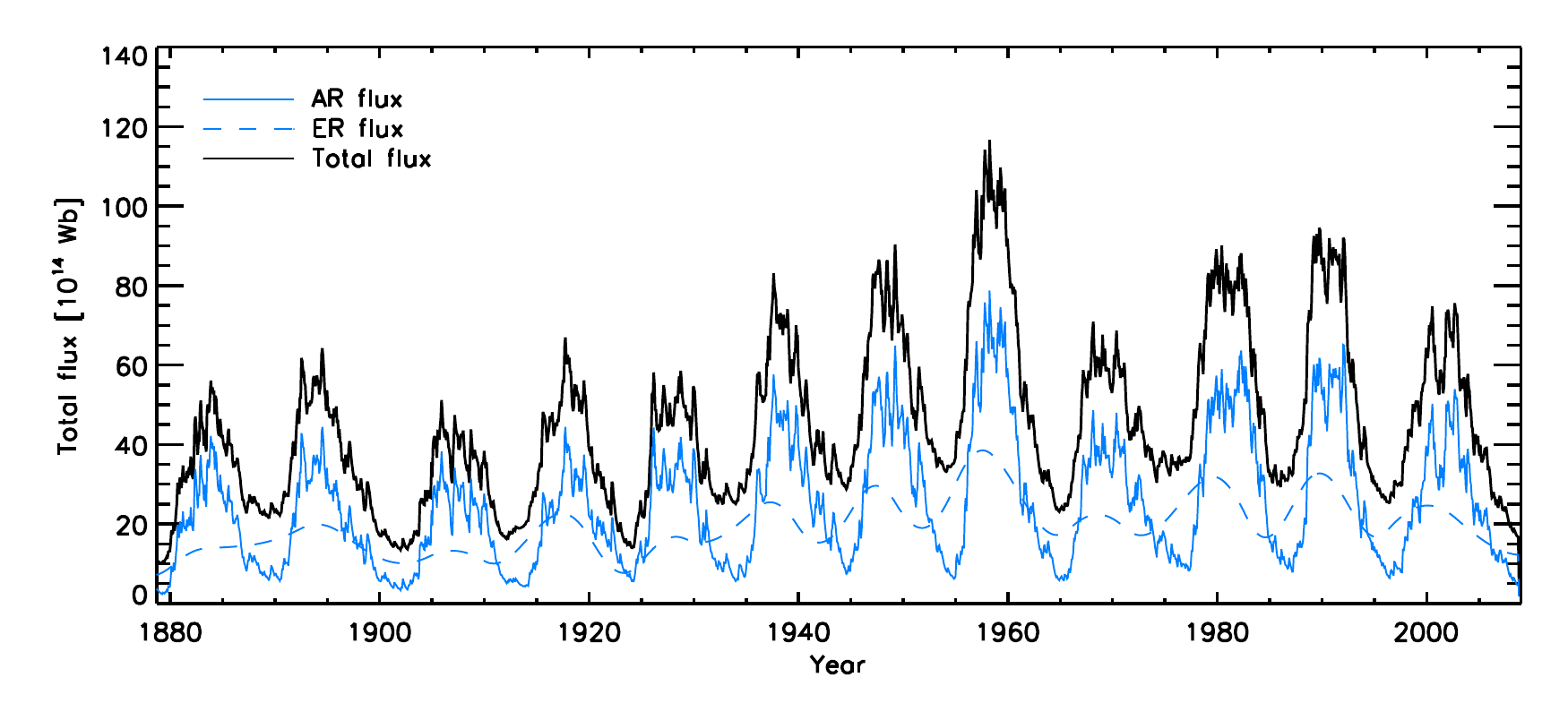}
  \caption{ The reconstructed total surface flux from 1878 to 2009 (black). The solid and dashed blue curves depict the contribution from active (AR) and ephemeral (ER) regions, respectively.}
 \label{totf_1874}
\end{figure*}

\begin{figure*}
   \centering
   \includegraphics[width=\textwidth]{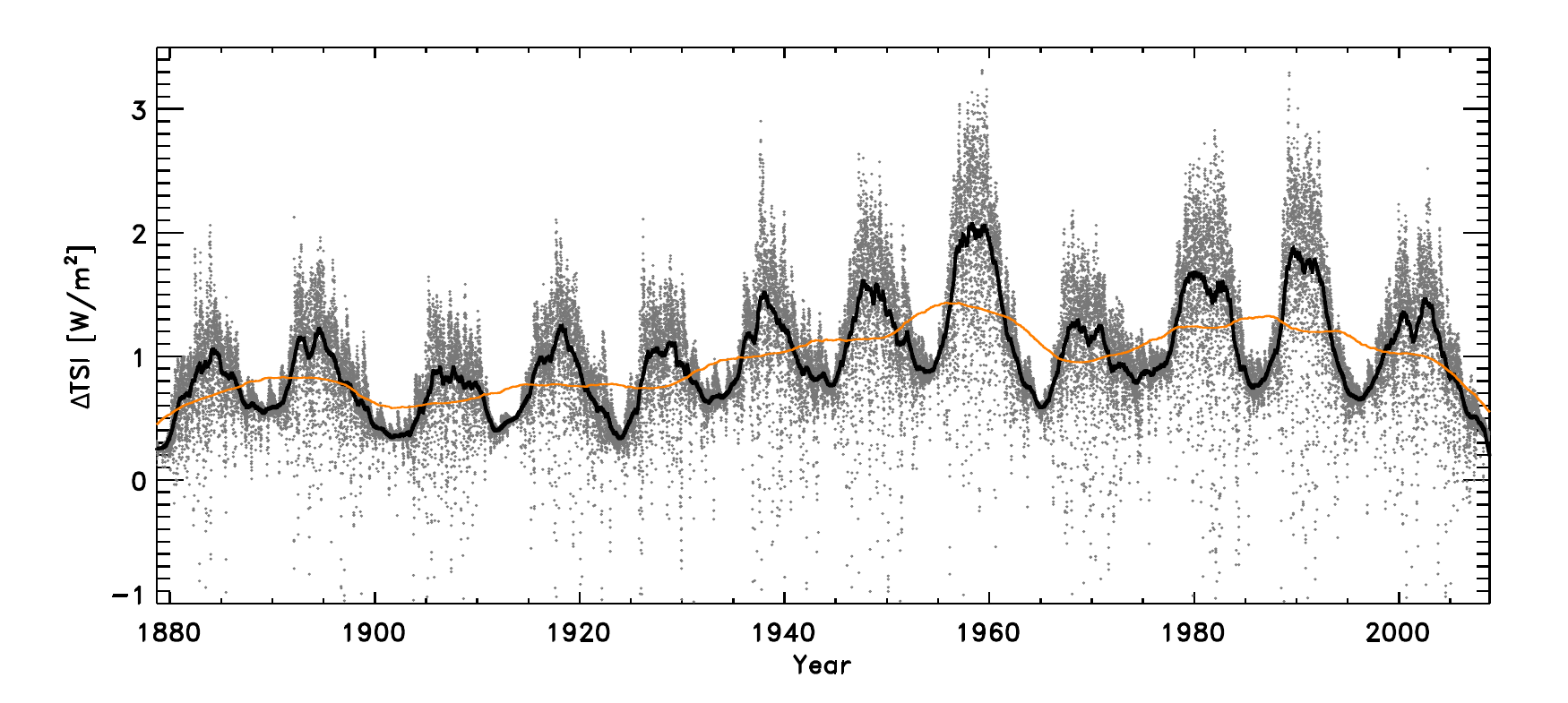}
         \caption{Change in TSI relative to the quiet Sun value, $\Delta \rm{TSI}$, between 1878 and 2009. Daily values are indicated by the light grey dots, and the yearly and 11-year smoothed values are depicted by black and orange curves, respectively.}
      \label{TSI_1878}
\end{figure*}

  \subsection{Total solar irradiance since 1878}

With all the free parameters fixed (as discussed in Sect. 4.1), we reconstruct the total magnetic flux and the TSI since 1878 (the beginning of cycle 12). The RGO-SOON sunspot group record starts in 1874, but the choice of the initial magnetic field configuration in the SFTM may influence the results of the first four to five years (i.e. the last years of cycle 11). 
Therefore, we show and discuss the reconstructions from 1878 onwards.
The reconstruction ends in 2009 (the end of cycle 23) since the TSI reconstruction of cycle 24 requires the amplitude and length of both cycles 24 and 25 in order to add the contribution of the ephemeral region flux.

The total magnetic flux since 1878 is plotted in Fig.~\ref{totf_1874} together with the corresponding contributions of active and ephemeral regions.
Ephemeral regions dominate during times of minima, when their magnetic flux surpasses that of the active regions \citep{harvey93, jin11}.
Although the flux at activity minima that originally emerged in the active regions shows a slight increase between the beginning and the second half of the 20th century, the main contribution to the change in the minima level in the total magnetic flux comes from the ephemeral regions.

\begin{figure}
   \centering
   \includegraphics[width=0.5\textwidth]{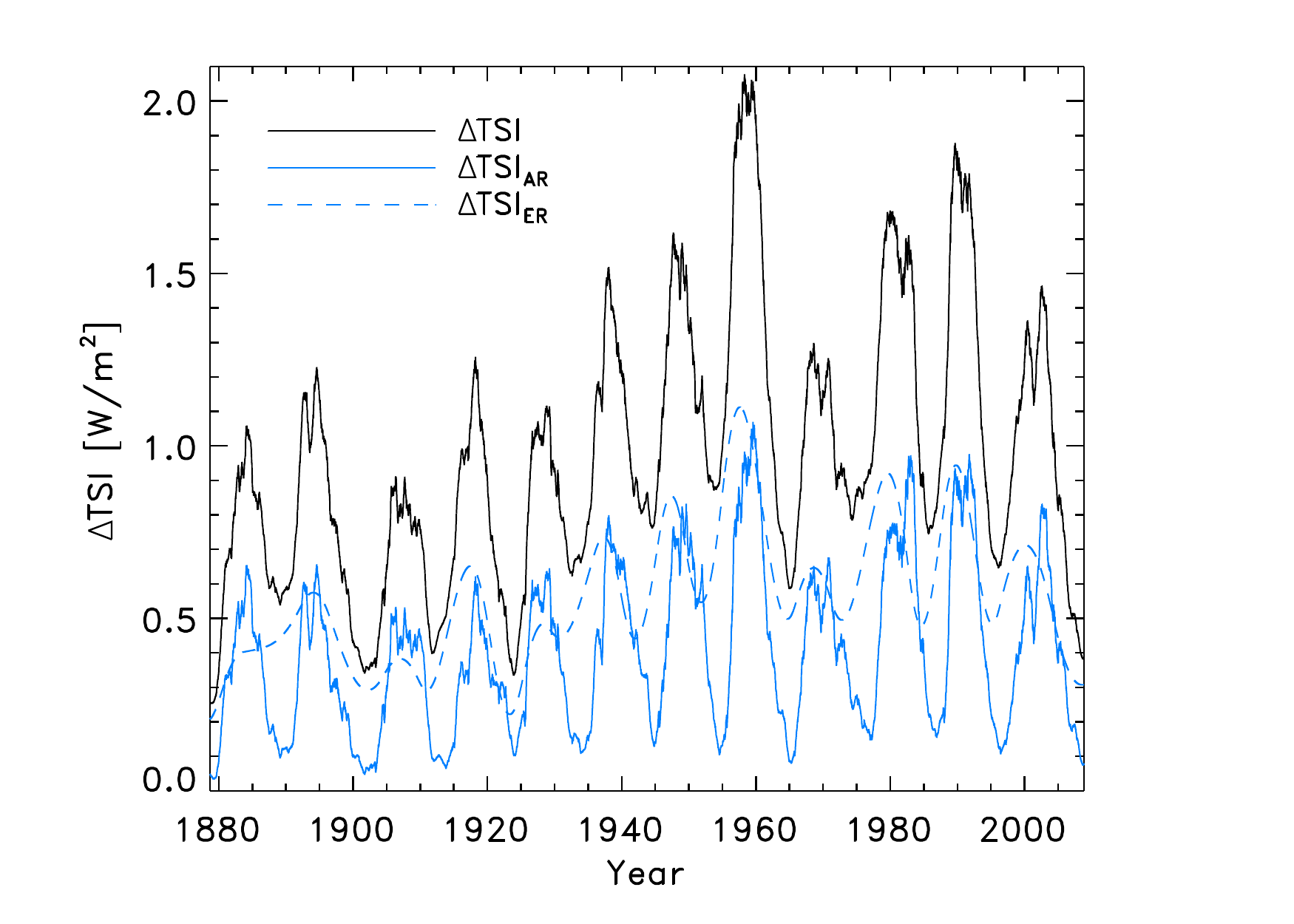}
         \caption{Change in TSI relative to the quiet Sun value smoothed over 1 year, $\Delta \rm{TSI}$ (black), as well as the contributions from active regions, $\Delta \rm{TSI}_{AR}$ (solid blue line), and ephemeral regions, $\Delta \rm{TSI}_{ER}$ (dashed blue line). }
      \label{DTSI_er_ar}
\end{figure}

The reconstructed TSI since 1878 is plotted in Fig.~\ref{TSI_1878} as the difference of the TSI to that from the entirely quiet Sun, i.e. $\Delta$TSI. The light grey dots correspond to the daily values, while the black and orange lines represent yearly and 11-year smoothed values, respectively.
In Fig.~\ref{DTSI_er_ar} we show the $\Delta$TSI smoothed over 1 year, together with the individual contributions from active (i.e. faculae and sunspots) and ephemeral regions.
At activity maxima, the contribution of active and ephemeral regions is comparable. However, as for the total magnetic flux, changes in the TSI minima levels are dominated by the ephemeral regions, which provide a larger secular variation than the active regions alone.

The ephemeral region cycles hold the largest uncertainties since little is known about the relationship between the length and amplitude of the ephemeral region cycles and those of the corresponding sunspot cycles, especially at earlier times.
We tested the results under different assumptions on the length and the amplitude of the ephemeral region cycles within the observational constraints \citep{harvey93, harvey01, hagenaar03, jin11}. We allowed the maxima of the ephemeral region cycles to be shifted (through a free parameter) between 2--5 years ahead of the associated active region maxima; the extension of the ephemeral region cycles $c_{x}$ to be inversely proportional to the length of the associated active region cycle; and combinations of these.
We found that the best-fit solution for the total magnetic flux predicts a very similar secular increase in all cases. In fact, the 11-year smoothed best fit curves of total magnetic flux since 1878 looked very similar to each other, irrespective of the exact recipe used for the ephemeral region cycle. The same happens with the secular increase in TSI since the TSI variation generally follows the evolution of the total magnetic flux.

Another source of uncertainty is the scatter of the 
tilt angle distribution. \cite{jiang14} show 
that the tilt angle scatter in the Mount Wilson and Kodaikanal data sets depends on the size of the sunspot group, ranging from $\sim$30$^{\circ}$ to $\sim$10$^{\circ}$ for the smallest and largest sunspot groups, respectively. When this scatter was introduced in an SFTM, it led to variations in the polar fields of the order of 30\% (standard deviation) for cycle 17, with particularly strong variations 
occurring when highly tilted sunspot groups emerged near the equator.
As a test, here we have calculated the change in TSI due to a variation in the tilt angle of a single active region (AR7999) with an area of 1232 $\mu$Hemispheres and at a latitude of -4$^{\circ}$, at the time of emergence.
In the reference case we used  the observed tilt angle, 1.89$^{\circ}$, and then increased it to a value of 50$^{\circ}$. After one year, the TSI was 0.08 W/m$^{2}$ higher compared to the reference case.
We also computed the TSI after a 40\% increase in the tilt angle of all sunspot groups in cycle 22, to test the combined effect. This produced an increase of $\sim$0.1 W/m$^{2}$ between the maximum of cycle 22 and the subsequent minimum.
The results of these simple tests demonstrate that the tilt angles have an effect on the modelled TSI, and therefore we plan to study how the scatter in tilt angles known from the observational data affects the TSI on solar cycle timescales in a subsequent study.

%__________________________________________________________________
%__________________________________________________________________

\section{Conclusions}

We have modelled the TSI starting from 1878 using an SFTM \citep{cameron10, jiang11} in combination with the SATIRE model.
The SFTM describes the evolution of the magnetic field on the solar surface due to active regions (sunspots and faculae) and their decay products based on the observed record of sunspot group areas and positions. The SFTM is used to simulate daily magnetograms of the photospheric magnetic field since 1878. The disc-integrated magnetic flux emerging in ephemeral regions, which is the prime contributor to the secular variation in the total photospheric flux and in the irradiance, was added separately following \cite{solanki02-b} and \cite{vieira10}.
The daily simulated magnetograms are then treated in a similar way as in the SATIRE-S model \citep{fligge00, krivova03, wenzler04} which relies on the observed magnetograms to calculate the filling factors of different photospheric components.
The model has been adapted to work with the low spatial resolution of the simulated magnetograms to obtain the contribution of the faculae. To derive the contribution of sunspots,  we used the record of sunspot group areas and positions.

The new model, SATIRE-T2 (for Telescope era version 2), has five free parameters that are fixed by comparing the simulated total photospheric flux and TSI to the observations available over the past 45 and 35 years, respectively.
The modelled TSI is in good agreement with the PMOD composite of observations, and reproduces daily and rotational variations more accurately than SATIRE-T.
Unfortunately, the period for which we have TSI measurements covers only the last three solar cycles, which were very similar in amplitude and activity.
Cycle 24 is turning out to be significantly weaker than the previous three cycles (comparable in sunspot number and area to cycle 14), so continuous TSI measurements until the end of cycle 24 will provide a crucial test to this and other models aiming to reconstruct TSI.

With the free parameters fixed to fit the observations, we then reconstructed the TSI since 1878.
The two main sources of uncertainty in our model are (1) the relationships between the parameters of the ephemeral region cycles (length and amplitude) and their corresponding sunspot cycles, and (2) the relationship between the cycle-average tilt angles and the strength of a cycle.
%Since the uncertainties in the relationship between the parameters of the ephemeral region cycles (length and amplitude) and their corresponding sunspot cycles are large, 
Therefore, we tested the sensitivity of the results to these relationships. We considered cases when the ephemeral region cycles peak several years before the associated sunspot cycle maximum, and when the extension in time of the ephemeral region cycles (relative to the sunspot cycle) is both proportional and inversely proportional to the length of the corresponding sunspot cycle.
In all of these cases the magnitude of the secular increase in the total magnetic flux and TSI since 1878 is very similar. 
%This testifies to the robustness of the general properties of our TSI reconstruction.
Two additional tests showed that the TSI is affected by variations in the tilt angle of the sunspot groups, which therefore indicates that including the observed scatter of the tilt angles will affect the TSI cycle amplitudes.

%We find that the TSI of the first two weak cycles (1880 -- 1900) is relatively high compared to the TSI reconstruction of the previous SATIRE-T model \citep{krivova10}.
%The relatively high TSI of these early cycles is attributed to the observed anti-correlation of the sunspot group tilt angles with the strength of a cycle \citep{Dasi-Espuig10, dasi-espuig13, kitchatinov11,mcclintock13}, as well as the correlation of the emergence latitudes of sunspot groups with cycle strength \citep{li03, solanki08, jiang11-a}, included in the SFTM used here.
%These two relationships introduce a modulation of the photospheric magnetic flux, especially low-order multipoles, that depends on the cycle strength.

Our irradiance reconstruction relies on the record of observed sunspot areas and positions, and for this reason it does not go farther back than 1874, i.e. 1878 after allowing the magnetic field in the activity belt in the SFTM to become independent of the initial condition. However, the model can be extended back to 1700 employing the reconstructed butterfly diagram from \cite{jiang11-a, jiang11}. The reconstructed butterfly diagram is a record of sunspot group areas and positions based on a statistical study of the sunspot number on the one hand, and the distributions of latitude, longitude, area, and tilt angle of sunspot groups per cycle and per cycle phase on the other.
In a subsequent paper we will study the total and spectral irradiance variations since 1700 using simulated magnetograms from the SFTM fed from the reconstructed record of sunspot group areas and positions. We will also include a detailed comparison with earlier reconstructions of solar irradiance, such as those by \cite{krivova10} and \cite{wang05}.

\begin{acknowledgements}
We thank Y-M. Wang for providing the composite of total magnetic flux measurements. M. D. E. would also like to thank L. Balmaceda for helpful discussions about PIKAIA.
The research leading to these results has received funding from the European Community's Seventh Framework Programme (FP7 2012) under grant agreement number 313188 (SOLID). J.J. acknowledges financial support from the National Natural Science Foundations of China through grant 11173033, and the Knowledge Innovation Program of the CAS (KJCX2-EW-T07).
The work of S. K. S. has been partly supported by the BK21plus program through the National Research Foundation (NRF) funded by the Ministry of Education of Korea.
\end{acknowledgements}

 \bibliographystyle{aa.bst}
 \bibliography{marybib.bib}

\end{document}